\documentclass[preprint,12pt]{elsarticle}

\usepackage{amssymb}
\journal{Physica A}

\begin{document}

\begin{frontmatter}



\title{Analysis of cross-correlations in electroencephalogram signals as an approach to proactive diagnosis of schizophrenia}


\author[1,2]{Serge F. Timashev}
\author[3]{Oleg Yu. Panischev}
\author[4]{Yuriy S. Polyakov\corref{cor1}}
\ead{ypolyakov@uspolyresearch.com}
\author[3]{Sergey A. Demin}
\author[5]{Alexander Ya. Kaplan}

\cortext[cor1]{Corresponding author at: USPolyResearch, 906 Spruce St., Ashland, PA, 17921. Tel.: 347-673-7747.}

\address[1]{Karpov Institute of Physical Chemistry, ul. Vorontsovo pole 10, Moscow, 103064 Russia}
\address[2]{Institute of Laser and Information Technologies, Russian Academy of Sciences, ul. Pionerskaya 2, Troitsk, 142092 Russia}
\address[3]{Institute of Physics, Kazan (Volga Region) Federal University, ul. Kremlevskaya 18, Kazan, 420008 Tatarstan, Russia}
\address[4]{USPolyResearch, 906 Spruce St., Ashland, PA 17921, U.S.A.}
\address[5]{Department of Human Physiology, Lomonosov Moscow State University, ul. Vorob'ovy Gory 1/12, Moscow, 119991 Russia}

\begin{abstract}
We apply flicker-noise spectroscopy (FNS), a time series analysis method operating on structure functions and power spectum estimates, to study the clinical electroencephalogram (EEG) signals recorded in children/adolescents (11 to 14 years of age) with diagnosed schizophrenia-spectrum symptoms at the National Center for Psychiatric Health (NCPH) of the Russian Academy of Medical Sciences. The EEG signals for these subjects were compared with the signals for a control sample of chronically depressed children/adolescents. The purpose of the study is to look for diagnostic signs of subjects' susceptibility to schizophrenia in the FNS parameters for specific electrodes and cross-correlations between the signals simultaneously measured at different points on the scalp. Our analysis of EEG signals from scalp-mounted electrodes at locations $F_3$ and $F_4$, which are symmetrically positioned in the left and right frontal areas of cerebral cortex, respectively, demonstrates an essential role of frequency-phase synchronization, a phenomenon representing specific correlations between the characteristic frequencies and phases of excitations in the brain. We introduce quantitative measures of frequency-phase synchronization and systematize the values of FNS parameters for the EEG data. The comparison of our results with the medical diagnoses for 84 subjects performed at NCPH makes it possible to group the EEG signals into 4 categories corresponding to different risk levels of subjects' susceptibility to schizophrenia. We suggest that the introduced quantitative characteristics and classification of cross-correlations may be used for the diagnosis of schizophrenia at the early stages of its development.

\end{abstract}

\begin{keyword}


Flicker-noise spectroscopy \sep Frequency-phase synchronization \sep Cross-correlations 
\sep Electroencephalogram signals \sep Schizophrenia
\end{keyword}

\end{frontmatter}

% \linenumbers

\section{Introduction}
\label{1}
The objective diagnosis of psychiatric disorders, such as schizophrenia, at the early stages of their development is complicated by the lack of reliable instrumental methods able to adequately describe these disorders at their onset \cite{Ris2010,Joh2009,Bas2008,Uhl2008}. For example; the methods of electro- and magnetoencephalography are rarely used for the diagnosis of psychiatric disorders, though there are experimental studies suggesting a high potential of these methods \cite{Kap2005,Gro2008,Sim2010}. This implies that clinical electroencephalogram (EEG) studies are of high concern for advancing the state of the art in the analysis, diagnosis, and prognosis of these disorders. 

Some aspects of the application of spectral methods to the analysis of EEG rhythmic activity in specific frequency ranges were previously discussed by \citet{Bor2005}. It was shown that one of the essential population signs of schizophrenia is the dampening of $\alpha$-activity accompanied by the amplification of low-frequency $\delta$- and $\theta$-activity. Segment analysis and statistical combinatorial analysis of spectral and segmental characteristics by pattern recognition were used to study the EEG temporal dynamics. The analysis demonstrated that the patients suffering from schizophrenia were characterized by reduced amplitudes and lengths and increased variabilities of quasistationary $\alpha$-activity segments, as compared to the controls. \citet{Kos2010} observed in patients suffering from schizophrenia an increase in the spectral density of EEG fluctuations in the low-frequency range (below 0.25 Hz). A significant role of stochastic components, specifically flicker noise, in EEG signals was discussed in depth by \citet{We08,All09,Aqu10}.

However, these conclusions are just empirical observations that cannot be clearly traced to the internal physiological mechanisms of psychiatric disorders. We believe there is an essential phenomenon of frequency-phase synchronization in the functional activity of some pathology-specific cortex areas that should be studied to provide a better understanding of the physiological aspects of schizophrenia as well as other neurodegenerative and psychiatric disorders. We expect that a cross-correlation analysis of frequency-phase synchronization will be more robust to the individual-specific variations of biomedical signals, which generally impede the conventional diagnosis. It should be noted that such an analysis cannot completely replace the characterization of specific EEG signals, but should rather complement it to develop a more comprehensive picture of the interactions between specific areas of the cortex.

Frequency and phase synchronization, manifestation of specific correlations between characteristic frequencies and phases of the excitations in different parts of the cortex (specific neural ensembles), and synchronization of the excitation amplitudes are the necessary conditions for the brain to function as an integral system \cite{Ros1996,Sin1993,Var2001,War2010}. A normally functioning brain responds to external actions on the human organism by establishing some optimal level of such synchronization. A significant deviation from this optimal level, such as an anomalously high level of synchronization or lack of synchronization, may be considered as an indicator of a pathology in brain activity. In the fields of medicine where digitized data, such as EEG \cite{Hor2006,Fer2009} or magnetoencephalograms (MEG) \cite{Bha2004,Yul2007}, are already recorded for the diagnosis, this optimal level can be estimated by analyzing the correlations between the signals recorded at spatially separated areas.

Studies of cooperative neural processes in bioelectric brain activity, which bring about frequency-phase synchronization and other similar phenomena, are relevant not only for the diagnosis of psychiatric disorders, but also for some neurodegenerative disorders, such as Parkinson's disease \cite{Tas1998} and temporal lobe epilepsy \cite{Nor2000}, and are important for the understanding of human cognitive abilities \cite{Bha2001}. The characteristic frequencies and relations between phase differences for EEG and MEG signals recorded at spatially separated electrodes are, as a rule, estimated using the Hilbert and Fourier transforms \cite{Ros1996,Ros2004} or wavelet analysis \cite{Lac1999}. Recently, the methods of cross-correlation analysis were also introduced into the analysis of mutual dynamics of physiological signals \cite{Pan2010,Pod2008,Pod10,Bob10}. \citet{Pan2010} applied the formalism of memory functions to study stimulated neuromagetic responses, MEG signals simultaneously recorded by different pairs of superconducting quantum interference device (SQUID) sensors located in spatially separated cortex areas. The analysis of synchronization effects in the mutual dynamics of neuromagnetic responses to flickering-color stimuli showed an increased level of frequency-phase synchronization in control (healthy) subjects compared to the patient suffering from photosensitive epilepsy. A detrended cross-correlation analysis (DCCA) method was proposed and applied to quantify cross-correlations in the presence of nonstationarity \cite{Pod2008,Pod09}. Another method, random matrix theory (RMT), which was developed to measure cross-correlations in collective modes for a large number of simultaneously recorded data \cite{Lal99,Ple99}, was employed to detect long-range power-law correlations in physiological data \cite{Pod10}. \citet{Bob10} reported that cross-correlations between pairs of EEG time series are inversely related to dissociative symptoms (psychometric measures) in 58 patients with paranoid schizophrenia.

In this study, we look for diagnostic signs of the susceptibility of children and adolescents to schizophrenia at its onset by analyzing the EEG signals recorded from scalp-mounted electrodes at locations $F_3$ and $F_4$. The clinical EEG signals were recorded in children/adolescents (11 to 14 years of age) with diagnosed schizophrenia-spectrum symptoms at the National Center for Psychiatric Health (NCPH) of the Russian Academy of Medical Sciences (Moscow, Russia). The EEG signals for these subjects were compared with the signals for a control sample of chronically depressed children/adolescents, also recorded at NCPH.

We perform the analysis using flicker-noise spectroscopy (FNS), a time series analysis method that introduces information parameters characterizing the components of stochastic signals in different frequency ranges and two-parameter cross-correlation functions \cite{Tim06a,Tim07a,Tim07b,Tim08a,Tim09,Tim10a,Tim10b}. The FNS method separates the analyzed signal into three components: low-frequency regular component corresponding to system-specific ``resonances" and their interferential 
 contributions, stochastic random-walk component at larger frequencies corresponding to dissipation effects, and highest-frequency inertial ``spike" component corresponding to flicker noise \cite [chap. 4.3] {Sch84}. It should be noted that the latter is taken into consideration due to the intermittent dynamics on various space-time scales attributed to complex (multiparticle, nonlinear) interactions, dissipation processes, and inertia characterizing thermodynamically open complex systems \cite{Bak97}.  

The FNS method was previously used to study frequency-phase synchronization between spatially separated cortex areas for the case of photosensitive epilepsy \cite{Tim09,Tim10b}. Neuromagnetic responses to external flickering-color stimuli, the prolonged action of which can cause epileptic seizures, were examined. The FNS analysis detected a disorder-specific feature, manifestation of high-frequency components (approximately 50 and 100 Hz) in the power spectum estimates for some SQUID-sensors. The analysis of FNS cross-correlation functions showed substantial disruptions in frequency-phase synchronization for the MEG dynamics of patient cortex areas compared to the synchronization level of control subjects \cite{Tim10b}.

The paper is structured as follows. In Section 2, we
provide the fundamentals of FNS and present the parameterization
algorithm and two-parameter cross-correlation function. In Section
3, we perform the parameterization of EEG signals at electrodes $F_3$ and $F_4$. In Section 4, we present the results of cross-correlation analysis of the signals, which suggest that the level of frequency-phase synchronization can be considered as a diagnostic sign of subjects' susceptibility to schizophrenia. In Section 5, we introduce quantitative measures of frequency-phase synchronization that allow us to partition all 84 subjects of the clinical study into 4 categories corresponding to different levels of the risk for developing schizophrenia. In Section 6, we discuss the potential capabilities of frequency-phase synchronization analysis for the diagnosis of neurological and psychiatric disorders.

\section{Principles of flicker-noise spectroscopy and basic relations}
\label{2}
 Here, we will only deal with the basic FNS relations needed to understand the parameterization procedure and cross-correlation function. FNS is described in more detail elsewhere \cite{Tim06a,Tim07a,Tim07b,Tim08a,Tim10a}.

In FNS, all introduced parameters for signal $V(t)$, where $t$ is time, are related to the autocorrelation function
\begin{equation}
\psi (\tau) = \left\langle {V(t)V(t + \tau )} \right\rangle_{T-\tau}, \label{eq1}
\end{equation}
where $\tau$ is the time lag parameter ($0 < \tau \le T_M$) and $T_M$ is the upper bound for $\tau$ ($T_M \le T/2$). This function characterizes the correlation in values of dynamic variable $V$  at higher, $t+\tau$, and lower, $t$, values of the argument. The angular brackets in relation (\ref{eq1}) stand for the averaging over time interval $T-\tau$:  
\begin{equation}
\left\langle {(...)} \right\rangle_{T-\tau}  = {1 \over {T-\tau}}\int^{T-\tau}_{0} {(...) \,dt}. \label{eq2}
\end{equation} 
The averaging over interval $T-\tau$ implies that all the characteristics that can be extracted by analyzing functions $\psi(\tau)$ should be regarded as the average values on this interval.  To extract the information contained in $\psi (\tau )$ ($\left\langle {V(t)} \right\rangle = 0$  is assumed), the following transforms, or ``projections", of this 
function are analyzed: cosine transforms (power spectrum estimates) $S(f)$, where $f$ is the frequency,
\begin{equation}
S(f) = 2 \int^{T_M}_{0} { \left\langle {V(t)V(t + t_1 )} \right\rangle_{T-\tau} \, \cos({2 \pi f t_1}) \,dt_1} \label{eq3}
\end{equation}
\newline and its difference moments (Kolmogorov transient structure functions) of the second order $\Phi^{(2)} (\tau)$
%eq4
\begin{equation}
\Phi^{(2)} (\tau) = \left\langle {\left[ {V(t) - V(t+\tau )} \right]^2 } \right\rangle_{T-\tau}. \label{eq4}
\end{equation}

The information contents of $S(f)$ and $\Phi^{(2)}(\tau)$ are generally different, and the parameters for both functions are needed to solve parameterization problems. By considering the intermittent character of signals under study, interpolation expressions for the stochastic components ${\Phi_s}^{(2)} (\tau)$ and $S_s(f)$ of $S(f)$ and $\Phi^{(2)} (\tau)$, respectively, were derived using the theory of generalized functions by \citet{Tim06a}. It was shown that the stochastic components of structure functions $\Phi^{(2)} (\tau)$ are formed only by jump-like irregularities (``random walks"), and stochastic components of functions $S(f)$, which characterize the ``energy side" of the process, are formed by spike-like (inertial) and jump-like irregularities.

\subsection{Signal parameterization}
\label{2.1}

In FNS parameterization, the original signal $V(t)$ is separated into three components: system-specific ``resonances" and their interferential 
 contributions at lower frequencies, stochastic jump-like (``random walk") component at larger frequencies, and stochastic spike-like (inertial) component in the highest 
frequency range. For simplicity, we will further refer to jump-like and spike-like irregularities as ``jumps" and ``spikes", respectively.

Let us write the basic interpolation expressions for stochastic components. The parameters characterizing the dynamic correlations on every level of the evolution hierarchy are assumed to be the same. Consider the simplest case, in which there is only one characteristic scale in the sequences of spikes and jumps \cite{Tim06a,Tim07a}:
\begin{equation}
\Phi_s ^{(2)} (\tau ) \approx 2\sigma ^2 \left[ {1 - \Gamma \,^{ - 1} (H_1 ) \cdot \Gamma \,(H_1 ,\tau /T_1 )} \right]^{_2 }, \label{eq5}
\end{equation}
where $\Gamma (s, x) = \int\limits_x^\infty  {\exp(-t) \cdot t^{s - 1}} dt$ and $\Gamma (s) = \Gamma (s, 0)$ are the complete and incomplete gamma functions, respectively ($x \ge 0$ and $s > 0$); $\sigma$ is the standard deviation of the measured dynamic variable with dimension [$V$]; $H_1$ is the Hurst constant, which describes the rate at which the dynamic variable ``forgets" its values on the time intervals that are less than the correlation time $T_1$.

For asymptotic cases, we obtain the formulas: 
\begin{equation}
\Phi_s^{(2)} (\tau ) = 2\Gamma ^{ - 2} (1 + H_1 ) \cdot \sigma ^2 \left( {{\tau  \over {T_1 }}} \right)^{2H_1 } ,\,\,\,\,\,\,\,{\tau  \over {T_1 }} \ll 1;
\label{eq6}
\end{equation}
\begin{equation}
\Phi _s^{(2)} (\tau ) = 2\sigma ^{2} \left[ {1 - \Gamma ^{ - 1} (H_1 ) \cdot \left( {{\tau  \over {T_1 }}} \right)^{H_1  - 1} \exp\left( { - {\tau  \over {T_1 }}} \right)} \right]^2 ,\,\,\,\,\,\,\,{\tau  \over {T_1 }} \gg 1.
\label{eq7}
\end{equation}

The interpolating function for power spectrum component $S_{sS}(f)$ formed by spikes can be written as: 
\begin{equation}
S_{sS} (f) \approx {{S_{sS} (0)} \over {1 + (2\pi f T_0 )^{n_0} }}.
\label{eq8}
\end{equation}
Here, $S_{sS}(0)$ is the parameter characterizing the low-frequency limit of $S_{sS}(f)$ and $n_0$ describes the degree of correlation loss in the sequence of spikes on the time interval $T_0$.

The interpolating function for the power spectrum component $S_{cJ}(f)$ formed by jumps is written as:
\begin{equation}
S_{sJ} (f) \approx {{S_{sJ} (0)} \over {1 + (2\pi  fT_1 )^{2H_1  + 1} }}.
\label{eq9}
\end{equation}
where $S_{sJ}(0)$ is the parameter characterizing the low-frequency limit of $S_{sJ}(f)$. 
	
Although the contributions to the overall power spectrum $S_s(f)$ given by Eqs. ~(\ref{eq8}) and ~(\ref{eq9}) are similar, the parameters in these equations can be much different: $S_{sJ}(0) \ne S_{sS}(0)$, $T_1 \ne T_0$, and $2H_1 + 1 \ne n_0$. This implies that the parameters in the expressions for the power spectrum and structure function generally have different information contents when the experimental time series $V(t)$ is analyzed. In cases when the contributions of spikes and jumps into the overall stochastic component are comparable, which corresponds to relatively small differences in the values of parameters $n_0$ and $2H_1+1$, it is convenient to use the following interpolating function for $S_s(f)$:
\begin{equation}
S_{s} (f) \approx {{S_{s} (0)} \over {1 + (2\pi f T_{01} )^{n} }}.
\label{eq10}
\end{equation}
where $S_s(0)$ and $n$ are phenomenological parameters.

The dynamics of complex systems includes both stochastic components, i.e., spikes and jumps, and system-specific slowly varying regular components associated with a set of frequencies. These frequencies correspond to internal and external resonances and their interferences. It should be noted that the whole set of resonance and interferential frequencies may get rearranged during the evolution of an open system. All the specific frequencies and their interferential contributions, which manifest themselves as oscillations in the dynamic variable $V(t)$, will be be further called ``resonant".

It is assumed that signal $V(t)$ can be presented as a linear superposition of stochastic component $V_s(t)$ and resonant component $V_r(t)$:
\begin{equation}
V(t) = V_s (t) + V_r (t).
\label{eq11}
\end{equation}

In this case, the autocorrelation function and power spectrum can be approximated as \cite{Tim07a,Tim08a,Tim09}:
\begin{equation}
\psi(\tau) = \psi_s (\tau) + \psi_r (\tau),
\label{eq12}
\end{equation}
\begin{equation}
S(f) = S_s (f) + S_r (f).
\label{eq13}
\end{equation}

Here, we assume that the resonant components are statistically stationary (they depend only on time lag $\tau$). This allows us to estimate $\psi_r (\tau)$ as an ``incomplete" cosine transform of $S_r (f)$ by applying the Wiener-Khinchin Theorem:
\begin{equation}
\psi _r (\tau ) \approx 2\int\limits_0^{f_{\max } } {S_r (f)\cos(2\pi f\tau )df},
\label{eq14}
\end{equation}
where $f_{\max }  = 0.5 f_d$, $f_d$ is the sampling frequency. It should be noted that Eq. (\ref{eq14}) is an approximation applied to a finite discrete time series assuming the wide-sense stationarity of the resonant signal component. The resonant component $\Phi_r^{(2)} (\tau)$ in this case is found by
\begin{equation}
\Phi _r^{(2)} (\tau ) = 2\left[ {\psi _r (0) - \psi _r (\tau )} \right].
\label{eq15}
\end{equation}

The stochastic component of $\Phi^{(2)} (\tau)$ can then be estimated as
\begin{equation}
\Phi_{s}^{(2)} (\tau ) \approx \Phi^{(2)} (\tau ) - \Phi _r^{(2)} (\tau ).
\label{eq16}
\end{equation}

Equations (\ref{eq12})-(\ref{eq16}) allow one to sequentally separate out resonant and stochastic components of structure functions and power spectrum estimates for experimental time series and perform the parameterization of the components, which is demonstrated in Refs. \cite{Tim07a,Tim10a}. The complete FNS parameterization algorithm in discrete form is listed in Appendix A.

In total, six stochastic FNS parameters are introduced ($T_{01}$, $S_s(T_{01}^{-1})$, $n$, $\sigma$, $T_1$, $H_1$): $T_{01}$, the correlation time for jump- and spike-like irregularities after which the 
 self-similarity observed in power spectrum estimate breaks down; $S_s(T_{01}^{-1})$, the ``spikiness" factor -- power spectrum estimate at frequency $T_0^{-1}$, which 
 accounts for the ``intensity" of jump- and spike-like irregularities in the highest-frequency interval; $n$, the flicker-noise parameter, 
 which characterizes the rate of loss of correlations in the series of high-frequency irregularities in time intervals $T_{01}$; $\sigma$, the 
 standard deviation of the value of the measured dynamic variable from the slowly varying resonant (regular) component, which is based solely on 
 jump-like irregularities; $T_1$, the correlation time for jump-like irregularities in stochastically varying signal $V(t)$; $H_1$, the 
  Hurst exponent (this estimate of the Hurst component is also referred to in literature as the Hausdorff exponent), which describes the rate at which the dynamic variable 
``forgets" its values on the time intervals that are less than $T_1$.

\subsection{Cross-correlation function}
\label{2.2}
The information about the dynamics of correlations in variables $V_i(t)$ and $V_j(t)$, measured at different points $i$ and $j$, can be extracted by analyzing the temporal variations of various cross-correlation functions. Here, we will use the simplest ``two-point" correlation expression characterizing the links between $V_i(t)$ and $V_j(t)$ \cite{Tim07b,Tim10b}:
\begin{equation}
q_{ij} (\tau,\theta _{ij} ) = \left\langle {\left[ {{{V_i (t) - V_i (t + \tau )} \over {\sqrt {\Phi _i^{\left( 2 \right)} \left( \tau  \right)} \,\,}}} \right] \left[ {{{V_j (t + \theta _{ij} ) - V_j (t + \theta _{ij}  + \tau )} \over {\sqrt {\Phi _j^{\left( 2 \right)} \left( \tau  \right)} }}} \right]} \right\rangle _{T - \tau  - \left| {\theta _{ij} } \right|},\label{eq17}
\end{equation}
where $\tau$  is the ``lag" time ($\tau > 0$), $\theta_{ij}$ is the ``time shift" parameter.

The cross-correlation expression $q_{ij} (\tau ;\theta _{ij} )$ is a function of temporal parameters $\tau$ and $\theta_{ij}$, which can be represented as a three-dimensional plot. Of most interest for the analysis are the intervals of $\tau$ and $\theta_{ij}$ where the cross-correlation function $q_{ij}$ approaches positive unity (maximum level of positive correlations) or negative unity (maximum level of negative correlations). The value of $\theta_{ij}$ corresponding to maximum values of cross-correlation $q_{ij} (\tau ;\theta _{ij} )$ characterizes the cause-and-effect relation (``flow direction") between signals $V_i(t)$ and $V_j(t)$. When $\theta_{ij}>0$, the flow moves from point $i$ to point $j$, when $\theta_{ij}<0$, from  $j$ to $i$. When the distance between points $i$ and $j$ is fixed, the value of $\theta_{ij}$ can be used to estimate the rate of information transfer between these two points.

The magnitude and behavior of the two-parameter expression (\ref{eq17}) may significantly depend on the value of selected averaging interval $T$ and upper-bound values of $\tau$ and $\theta_{ij}$, which we will refer to as $\tau_{\max}$ and $\theta_{\max}$. From the statistical reliability point of view, we set a constraint of $\tau_{\max} + |\theta_{\max}| \le T/2$. 

For conciseness, from now on we will refer to $\theta_{ij}$ as $\theta$.

\section{Parameterization of EEG signals at $F_3$ and $F_4$ electrodes}
\label{3}
The EEG signals were recorded using Ag/AgCl electrodes (BrainAmp, Brain Products GbmH, Germany) in wakeful relaxed children/adolescents with the eyes closed at 16 electrode sites ($O_1$, $O_2$, $P_3$, $P_4$, $P_z$, $T_5$, $T_6$, $C_3$, $C_4$, $C_z$, $T_3$, $T_4$, $F_3$, $F_4$, $F_7$, and $F_8$) set according to the international 10-20 system. The participants were seated comfortably in a dimly lit, electrically shielded room. The electrodes were monopolarly referenced to coupled ear electrodes. Electrode impedances were monitored and kept below 10 K$\Omega$ during the entire experiment. The EEG signals were continuously recorded at a sampling rate of 256 Hz and then down-sampled off-line to 128 Hz ($f_d$ = 128 Hz). Only artifact-free EEG segments were used for the analysis. For every electrode, 7680 values of the electric potential were examined. 

Several longitudinal studies of schizophrenia suggest the existence of a cerebral degenerative process in, at least, some patients \cite{Del97,Gur98,Nair97}. According to studies \cite{Gur98,Rap99,Heu11}, this neurodegeneration may be most pronounced in the frontal lobes of the cerebrum. For this reason, we chose frontal electrodes $F_3$ and $F_4$ for our analysis. The examined signals $V_{F3}(t)$ and $V_{F4}(t)$ for electrodes $F_3$ and $F_4$, respectively, were recorded in 39 healthy subjects (marked with prefix ``S") and 45 subjects susceptible to schizophrenia. This initial grouping was done based on the medical diagnosis by specialists, which was performed at NCPH. Our analysis included the determination of subject-specific FNS parameters for the $F_3$ and $F_4$ signals and examination of cross-correlations between the signals by Eq. (\ref{eq17}). This comparative study allowed us to partition the EEG signals of 84 subjects into 4 categories corresponding to different levels of risk for developing schizophrenia, starting with group I (lowest risk) and ending with group 4 (highest risk). The subjects belonging to the first two groups may be characterized according to the medical diagnosis as ``healthy" while the last two groups correspond to ``susceptible to schizophrenia". For conciseness, in this and next sections we discuss in detail only selected examples of the analysis to illustrate our partitioning logic. The complete classification of all subjects is presented in Section 5.

First, we would like to note that the FNS parameterization alone of EEG signals generated in the cortex areas under electrodes $F_3$ and $F_4$ was insufficient to partition all 84 subjects into risk groups due to a high individual-specific variability of FNS parameters for the stochastic components of the signals. To unambiguously partition all subjects into 4 risk groups, we had to complement the FNS parameterization of $F_3$ and $F_4$ signals with their cross-correlation analysis using expression (\ref{eq17}).

\begin{figure}
\caption{\label{fig1} (Color online) EEG signals at electrode $F_3$ for typical members of groups I (a), II (b), III (c), and IV (d).}
\begin{center}
\includegraphics[width=14.5cm]{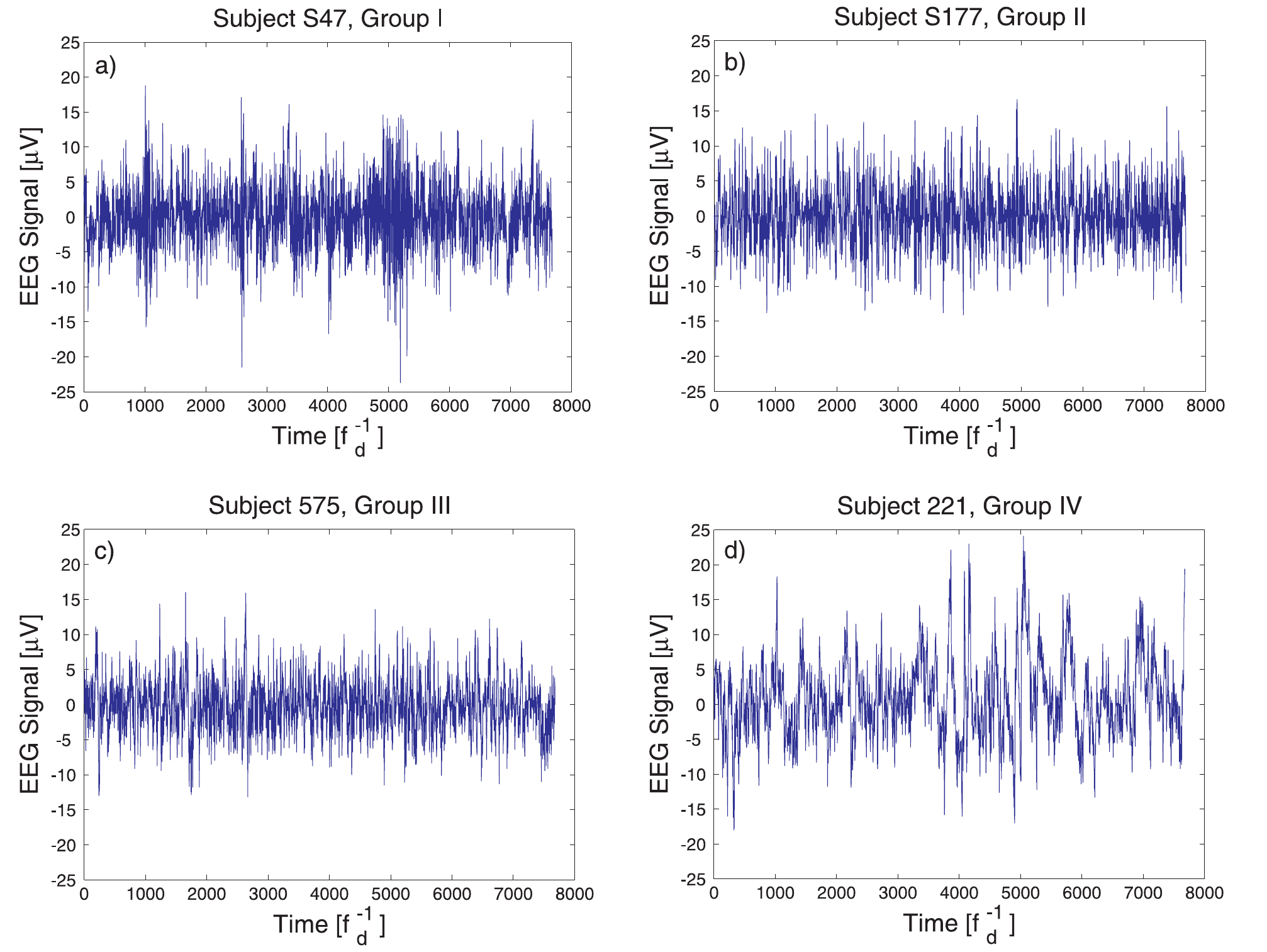}
\end{center}
\end{figure}

\begin{figure}
\caption{\label{fig2} (Color online) Linear-scale power spectrum estimates of EEG signals in Fig. \ref{fig1} for typical members of groups I (a), II (b), III (c), and IV (d).}
\begin{center}
\includegraphics[width=14.5cm]{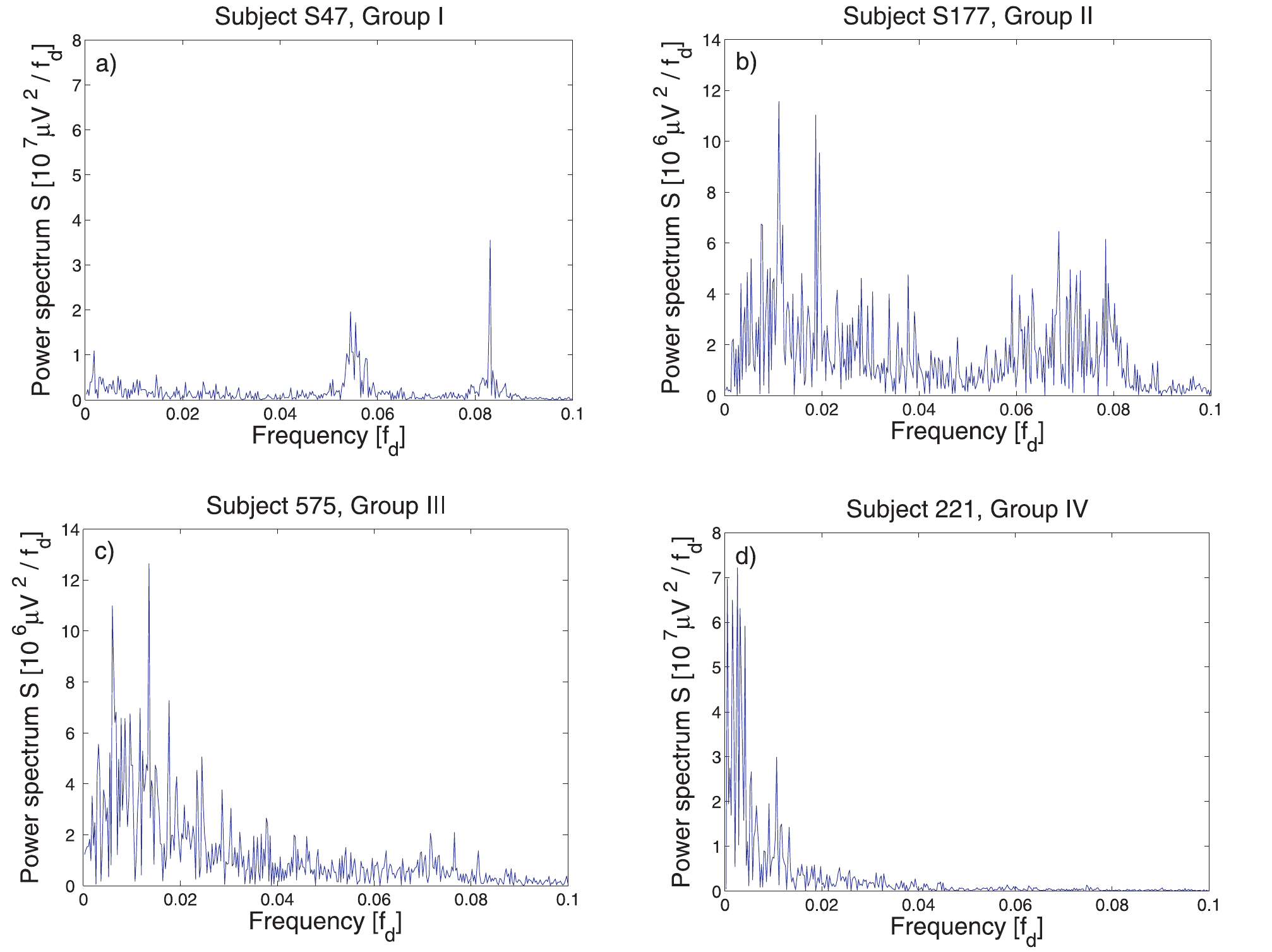}
\end{center}
\end{figure}

\begin{figure}
\caption{\label{fig3} (Color online) Log-log-scale power spectrum estimates of EEG signals in Fig. \ref{fig1} for typical members of groups I (a), II (b), III (c), and IV (d); 1 (blue) -- power spectrum estimate for the experimental data, 2 (red) -- stochastic interpolation $S_s(f)$.}
\begin{center}
\includegraphics[width=14.5cm]{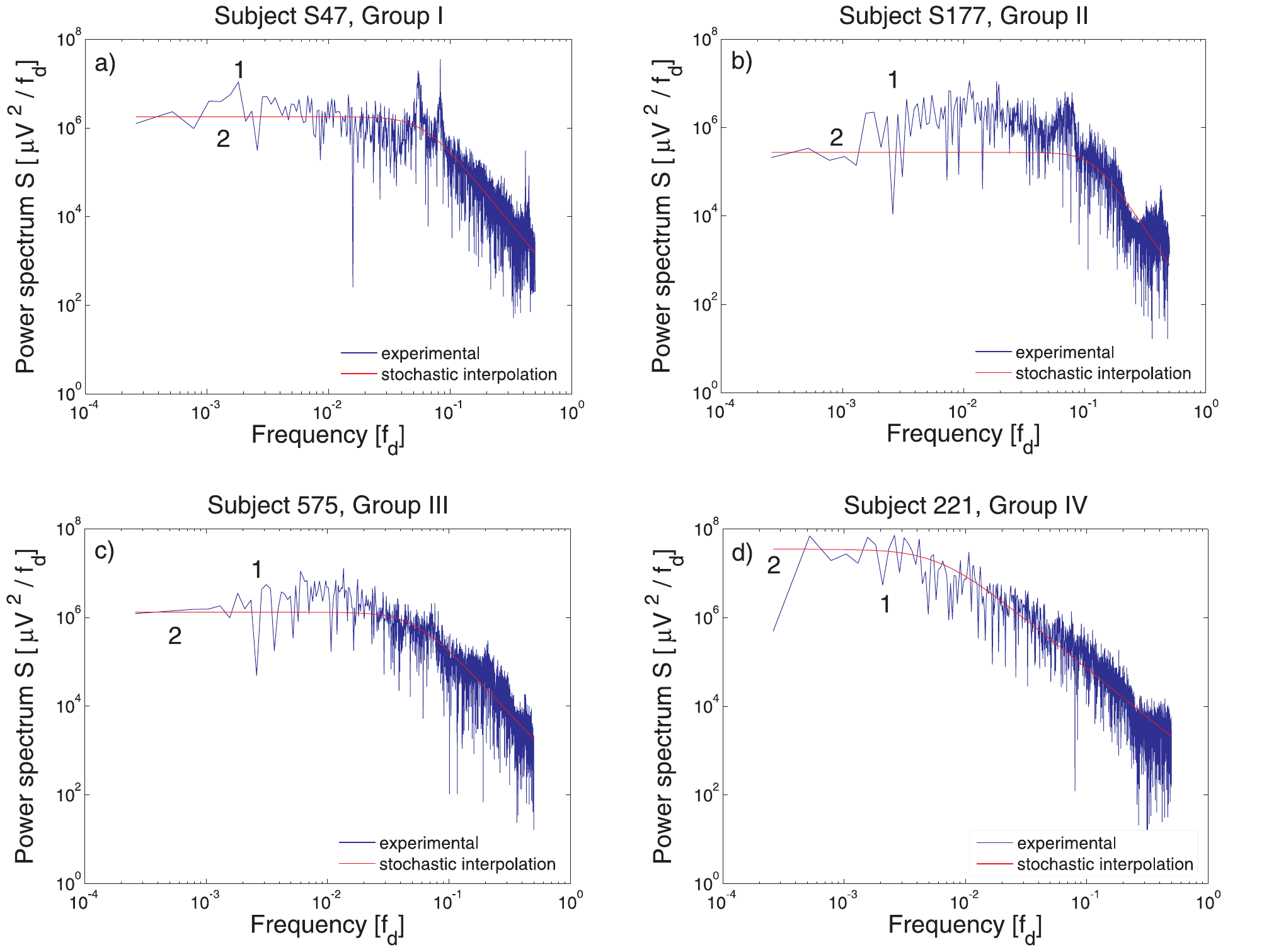}
\end{center}
\end{figure}

\begin{figure}
\caption{\label{fig4} (Color online) Structure functions of EEG signals in Fig. \ref{fig1} for typical members of groups I (a), II (b), III (c), and IV (d); 1 (blue) -- structure function for the experimental data, 2 (red) -- resonant component of the structure function, ``cross" (green) -- total FNS interpolation.}
\begin{center}
\includegraphics[width=14.5cm]{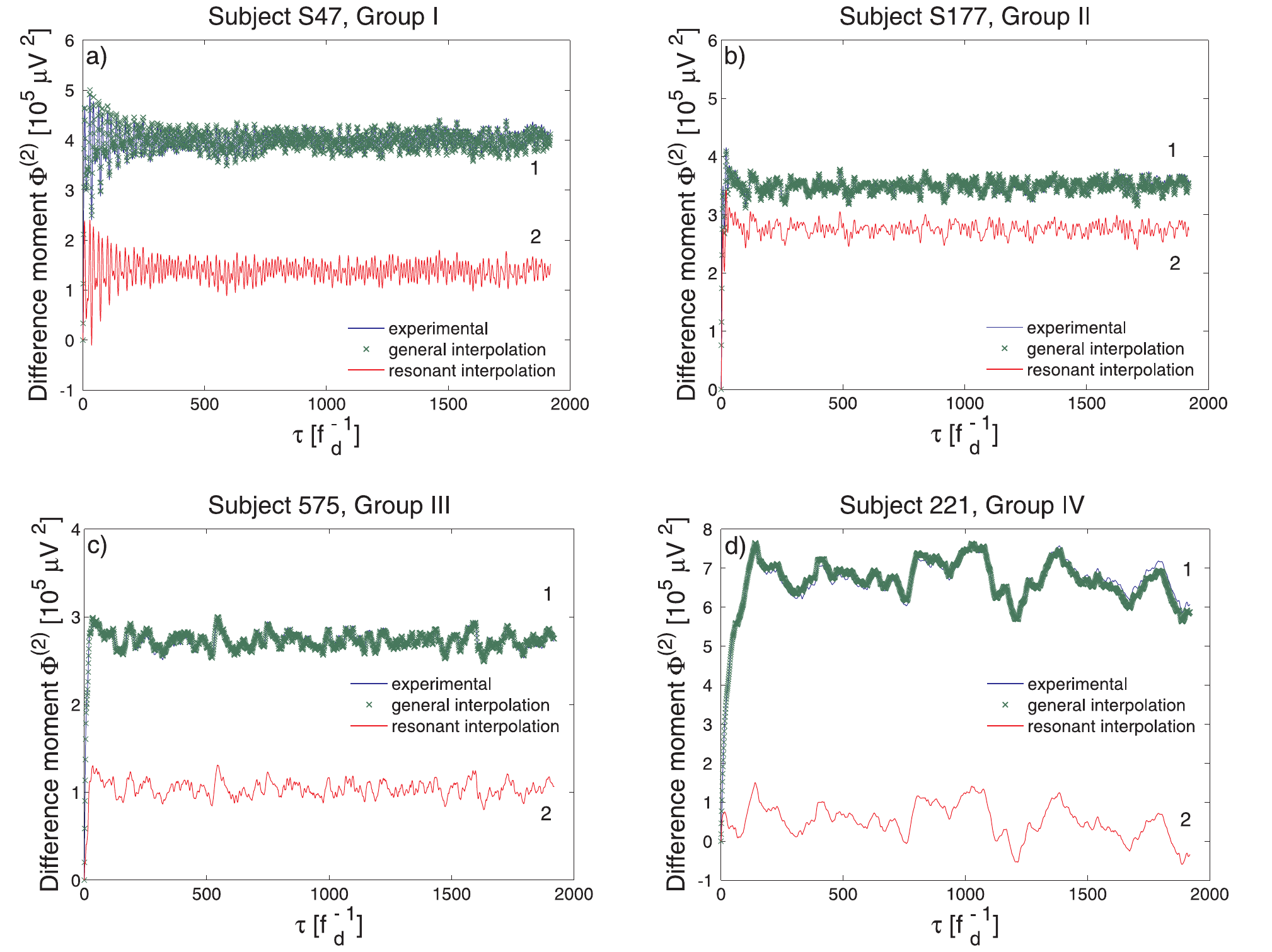}
\end{center}
\end{figure}

The degree of scatter in the values of FNS parameters can be seen from Table~\ref{tab1}, which shows the values of all FNS parameters for typical subjects of each of the four groups. The FNS parameterization was run using the algorithm listed in Appendix A for the averaging interval $T$ equal to the series length (7680 points). Some typical dependences used in FNS parameterization are illustrated in Figs. ~\ref{fig1}-\ref{fig4}. The complete partitioning of all subjects into four groups corresponding to different risk levels for developing schizophrena together with the values of group-determining FNS parameters and quantitative estimates of cross-correlations is presented in Table~\ref{tab2}.  Table~\ref{tab1} shows that the values of FNS parameters for subjects assigned to different groups may significantly overlap, which implies that FNS parameterization alone cannot be used for the classification. The combined analysis of FNS parameters and cross-correlations discussed in Sections 4 and 5 allowed us to suggest the spikiness factor $S_s(T_{01}^{-1})$ as the most functionally significant FNS parameter for the classification. Other FNS parameters listed in Table~\ref{tab1} have a rather limited value for the partitioning: it is possible to draw conclusions only on the form of their variability for subjects of different categories.

\begin{table}
\caption{\label{tab1}FNS parameters for typical members of each group. (*) denotes the cases of strong nonstationarity in EEG signals}
\scriptsize
\begin{center}
%  \begin{tabular}{|p{1.2cm}|p{1.2cm}|p{1.2cm}|p{1.2cm}|p{1.2cm}|p{2cm}|p{1.2cm}|p{1.2cm}|}
  \begin{tabular}{|c|c|c|c|c|c|c|c|}
    \hline
    Subject & Electr. & $\sigma,$ & $H_1$ & $T_1, \ f_d^{-1}$ & $S_s(T_{01}^{-1}),$ & $n$ & $T_{01}, \ f_d^{-1}$ \\
    $\mathstrut$  & $\mathstrut$ & $\mu$V & $\mathstrut$ & $\mathstrut$ & $\mu$V$^2 f_d^{-1}$ & $\mathstrut$ & $\mathstrut$ \\
     \hline \hline
     \multicolumn{8}{|c|}{Group I} \\ \hline
     \textbf{S47} & $F_3$ & 360 & 1.22 & 2.30 & $3.8\times 10^3$ & 3.36 & 2.64 \\ \cline{2-8}
      $\mathstrut$ & $F_4$ & 324 & 1.04 & 2.45 & $5.7\times10^3$ & 2.98 & 2.58 \\ \hline
     
     \textbf{S163} & $F_3$ &145 & 1.87 & 0.83 & $42.8$ & 4.45 & 1.30 \\ \cline{2-8}
     $\mathstrut$ & $F_4$ & 158 & 2.03 & 0.86 & 61.8 & 4.43 & 1.48 \\ \hline
     
     \textbf{S165} & $F_3$ & 293 & 1.84 & 1.46 & 801.26 & 3.96 & 2.39 \\ \cline{2-8}
     $\mathstrut$ & $F_4$ & 263 & 1.68 & 1.44 & 762.35 & 3.81 & 2.19 \\ \hline \hline

     \multicolumn{8}{|c|}{Group II} \\ \hline
     
     \textbf{S177*} & $F_3$ & 209 & 0.02 & $3.2\times10^5$ & 157 & 4.10 & 1.30 \\ \cline{2-8}
     $\mathstrut$ & $F_4$ & 170 & 0.05 & $9.1\times10^5$ & 24 & 4.35 & 0.87 \\ \hline
     
     \textbf{156} & $F_3$ & 282 & 1.13 & 4.64 & $2.42\times10^4$ & 2.58 & 6.85 \\ \cline{2-8}
     $\mathstrut$ & $F_4$ & 256 & 1.01 & 5.61 & $2.7\times10^4$ & 2.46 & 7.68 \\ \hline
     
     \textbf{S42} & $F_3$ & 196 & 2.24 & 0.57 & 26.27 & 4.95 & 1.04 \\ \cline{2-8}
     $\mathstrut$ & $F_4$ & 151 & 2.77 & 0.37 & 12.68 & 4.98 & 0.86 \\ \hline \hline

     \multicolumn{8}{|c|}{Group III} \\ \hline

    \textbf{575} & $F_3$ & 291 & 1.10 & 2.60 & $6.2\times10^3$ & 2.90 & 3.00 \\ \cline{2-8}
     $\mathstrut$ & $F_4$ & 235 & 1.31 & 1.86 & $2.4\times10^3$ & 3.11 & 2.51 \\ \hline 

     \textbf{S31} & $F_3$ & 382 & 1.12 & 3.24 & $1.5\times10^4$ & 2.89 & 3.95 \\ \cline{2-8}
     $\mathstrut$ & $F_4$ & 415 & 0.85 & 5.45 & $2.7\times10^4$ & 2.73 & 4.96 \\ \hline 

     \textbf{S12} & $F_3$ & 133 & 1.95 & 0.75 & 36.00 & 4.43 & 1.23 \\ \cline{2-8}
     $\mathstrut$ & $F_4$ & 118 & 3.02 & 0.38 & 32.55 & 4.45 & 1.18 \\ \hline \hline

     \multicolumn{8}{|c|}{Group IV} \\ \hline
     
     \textbf{221} & $F_3$ & 558 & 0.56 & 35.9 & $6.04\times10^5$ & 2.20 & 26.5 \\ \cline{2-8}
     $\mathstrut$ & $F_4$ & 520 & 0.56 & 37.9 & $6.58\times10^5$ & 2.10 & 29.8 \\ \hline

 \textbf{387\_03} & $F_3$ & 356 & 1.28 & 3.65 & $1.2\times10^4$ & 3.07 & 4.89 \\ \cline{2-8}
     $\mathstrut$ & $F_4$ & 311 & 1.35 & 3.21 & $7.8\times10^3$ & 3.11 & 4.56 \\ \hline

\textbf{573} & $F_3$ & 232 & 0.33 & 37.15 & $1.7\times10^5$ & 1.72 & 22.7 \\ \cline{2-8}
  $\mathstrut$ & $F_4$ & 228 & 0.46 & 12.25 & $5.9\times10^4$ & 1.87 & 9.51 \\ \hline

 \end{tabular}
\normalsize
\end{center}
\end{table}

The first two groups are denoted as ``healthy" (group I) and ``conditionally healthy" (group II). It can be seen that the EEG signals of groups I and II are characterized by a relatively rapid loss of correlations in stochastic components, which follows from high values of the Hurst component ($H_1 > 1$) and flicker-noise parameter $n$, and small correlation times $T_1$ and $T_{01}$ (Figs. \ref{fig3}a, \ref{fig3}b). The group of ``conditionally healthy" subjects also includes 5 subjects (S59, S170, S177, 351, S43) with a strong nonstationarity in the EEG signals: the steady-state variance is not reached within the analyzed interval (formally, this corresponds to $H_1 \ll 1$ and $T_1 \gg T$). The subjects in groups I and II are also characterized by moderate values of spikiness factor $S_s(T_{01}^{-1}) < 3 \times 10^4 \mu V^2 f_d^{-1}$. The third and fourth groups are composed of subjects ``susceptible to schizophrenia" and ``highly susceptible to schizophrenia", respectively. The EEG signals for these categories are characterized by a high degree of correlations in the stochastic irregularities of both types, which can be seen from the lower values of $H_1$ and $n$ as compared to groups I and II, and high values of spikiness factor $S_s(T_{01}^{-1})$ (up to $ 7 \times 10^5 \mu V^2 f_d^{-1})$. The correlation times $T_1$ and $T_{01}$ are small (Figs. \ref{fig3}c, \ref{fig3}d) in this case as well.

A significant scatter in the values of characteristic parameters was also observed for low-frequency spectra of the signals, i.e., in the peaks corresponding to a series of ``resonances". These ``resonance" frequencies can be attributed to individual-specific resonances and interferential effects caused by the interaction of intrinsic and external frequencies. The power spectrum estimates of EEG signals recorded by electrodes $F_3$ and $F_4$ for the subjects of group I (Fig.~\ref{fig1}a) display clear peaks in the frequency range from 5 to 12 Hz (Fig.~\ref{fig2}a). These peaks can be attributed to the brain $\alpha$-rythm, which manifests itself at the times of rest, relaxation, or slight meditation with the eyes closed. The dynamics of the difference moment for these EEG signals has a small-scale structure (Fig.~\ref{fig4}a).  The power spectrum estimates of EEG signals for the subjects of group II  (Fig.~\ref{fig1}b) contain more noise on the background of which there are peaks corresponding to brain rythms (Fig.~\ref{fig2}b). As compared to group I, the magnitude of the peaks is reduced and the peaks themselves are shifted to lower frequencies. These results are in agreement with prior results reported by \citet{Bor2005}, where the dampening of $\alpha$-activity and amplification of low-frequency $\delta$- and $\theta$-activity is considered as a diagnostic sign of schizophrenia. The scale of fluctuations in the difference moment is increased compared to the group I case  (Fig.~\ref{fig4}b). The EEG signals for the group III subjects (Fig.~\ref{fig1}c) are characterized by further reduction of the activity in the range from 5 to 12 Hz and higher peaks in the range of ultralow frequencies ($0.4 - 4$ Hz). The scale of fluctuations in the difference moment gets further increased (Fig.~\ref{fig4}c). In the group IV subjects (Fig.~\ref{fig1}d), the ultralow frequency dynamics becomes even more pronounced: peaks in the frequency range higher than 2.5 Hz have no noticeable effect and the large-magnitude peaks in the ultralow frequency range play the dominant role (Fig.~\ref{fig2}d). This observation is in agreement with the results of study \cite{Kos2010}, where it was noticed that the subjects suffering from schizophrenia have an increased spectral density of EEG signal fluctuations in the low-frequency range. In the group IV subjects, the structure of the difference moment is dominated by large-scale fluctuations  (Fig.~\ref{fig4}d).

\section{Frequency-phase synchronization in EEG signals at $F_3$ and $F_4$ electrodes}
\label{4}

We analyzed the dynamics of cross-correlations between EEG signals at electrodes $F_3$ and $F_4$ using Eq. (\ref{eq17}). The two-parameter cross-correlation dependences were calculated for adjacent intervals with length $T = 800 f_d^{-1} = 6.25$ s for all 84 subjects. This implies that the studied time interval 7200 $f_d^{-1}$ = 56.25 s was partitioned into 9 subintervals (1...800) $f_d^{-1}$ ,..., (6401...7200) $f_d^{-1}$, for each of which
 two- and three-dimensional plots of the cross-correlation function were built. The ranges of the input parameters were $0 \le \tau \le \tau_{\max} = 0.1 T$ and $-\theta_{\max} \le \theta \le \theta_{\max}$, where $\theta_{\max} = 0.4 T$. Here, we present plots only for selected subintervals to illustrate the general structure of observed cross-correlation variations.

\begin{figure}
\caption{\label{fig5} Typical 3-D plots of cross-correlation function $q(\tau,\theta)$ for EEG signals at locations $F_3$ and $F_4$ and their cross-sections at $\tau_0=40 f_d^{-1}$ built for subject S47, group I (a, c) and S177, group II (b, d).}
\begin{center}
\includegraphics[width=13.5cm]{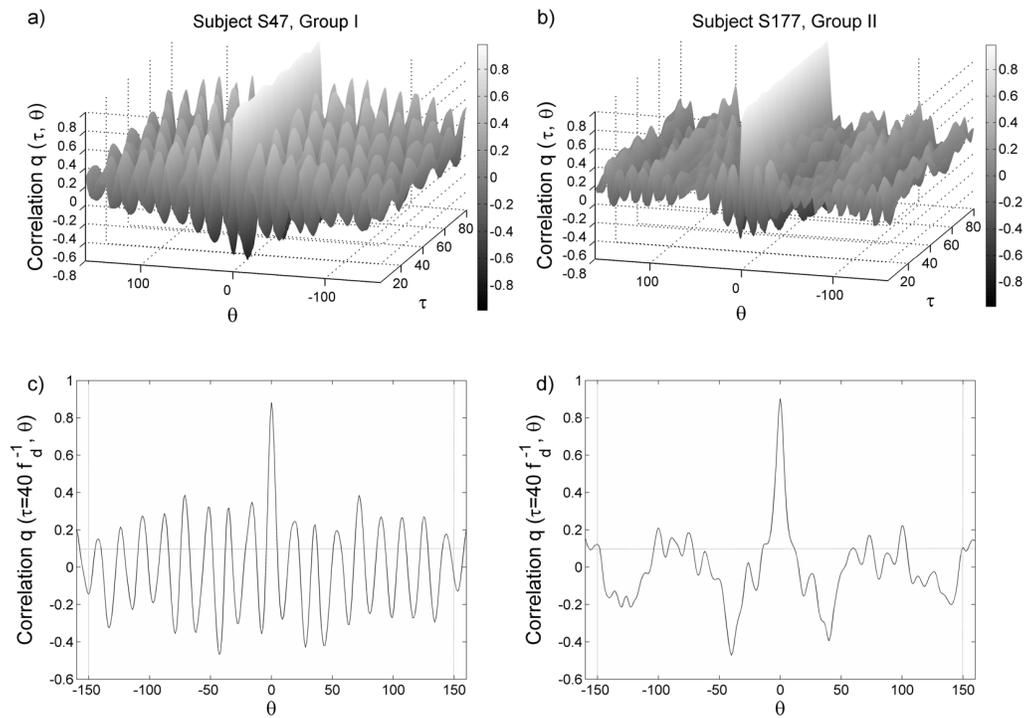}
\end{center}
\end{figure}

\begin{figure}
\caption{\label{fig6} Typical 3-D plots of cross-correlation function $q(\tau,\theta)$ for EEG signals at locations $F_3$ and $F_4$ and their cross-sections at $\tau_0=40 f_d^{-1}$ built for subject 575, group III (a, c) and 221, group IV (b, d).}
\begin{center}
\includegraphics[width=13.5cm]{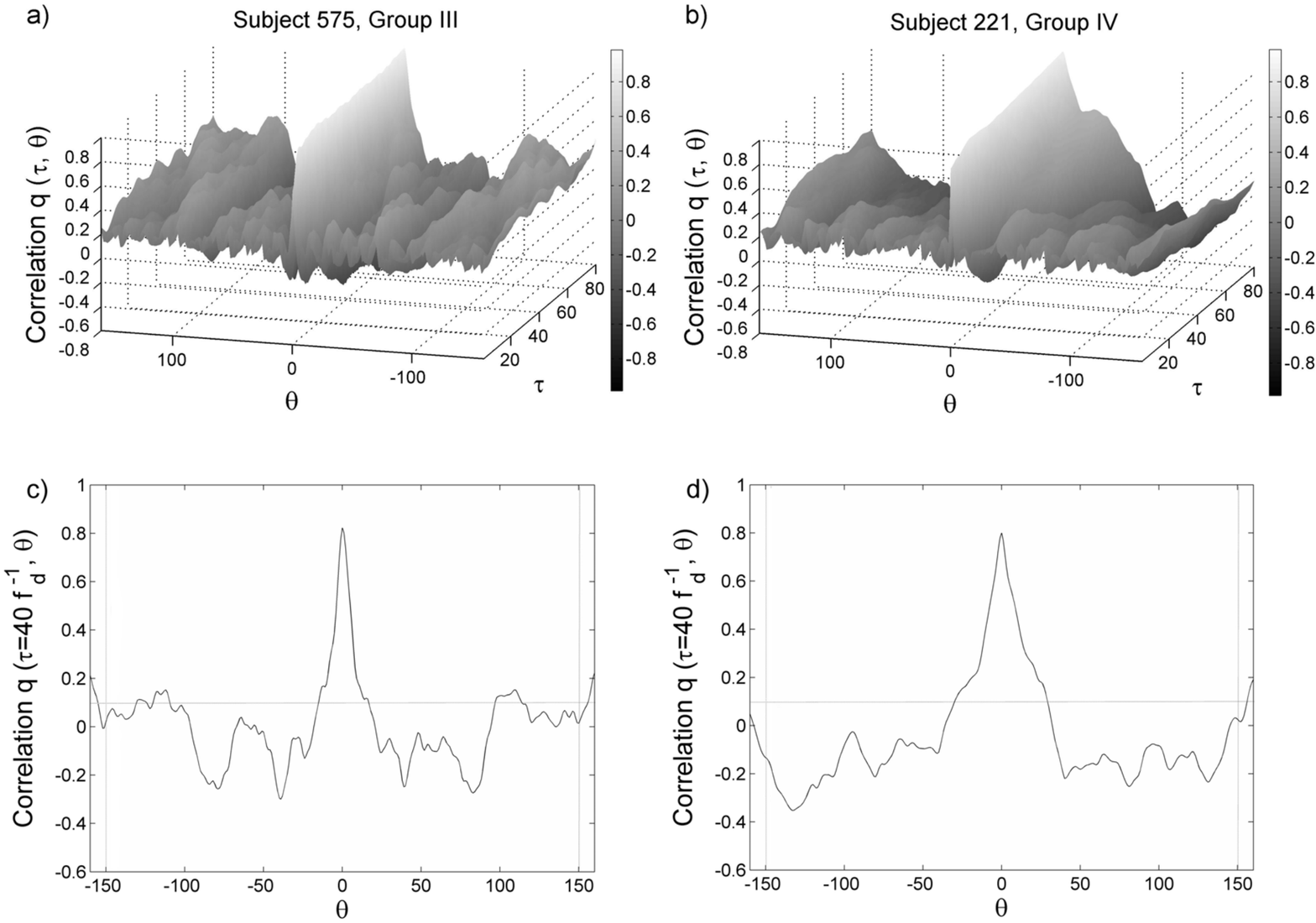}
\end{center}
\end{figure}

Figures \ref{fig5}-\ref{fig6} show three-dimensional plots of cross-correlation function $q(\tau,\theta)$ for the $F_3$ and $F_4$ EEG signals of some of the typical subjects that were considered in Section 3. These examples support our working hypothesis stating that the degree of disruption in frequency-phase synchronization within specific cortex areas (in this case, left and right frontal areas) can be used as a diagnostic sign of subjects' susceptibility to schizophrenia-spectrum disorders.

Figure \ref{fig5} illustrates the cross-correlation function and its plane projections for the EEG signals of subjects S47 (Figs. \ref{fig5}a, \ref{fig5}c) and S177  (Figs. \ref{fig5}b, \ref{fig5}d) built for one of the subintervals of length  800 $f_d^{-1}$. Frequency-phase synchronization between the signals $F_3$ and $F_4$ recorded in subject S47 manifests itself as a periodic dependence of  $q(\tau,\theta)$ on $\theta$, which can be best illustrated by looking at the cross-section of the three-dimensional plot at $\tau_0 =$ const. Figure \ref{fig5}c shows the cross-section at $\tau_0 =40 f_d^{-1}$ (this choice of cross-section will be explained later). The maximum value of cross-correlation function $q(40 f_d^{-1},0) \approx 0.89$ suggests a high level of correlation between simultaneously recorded $V_{F3}$ and $V_{F4}$. A local maximum $q(40 f_d^{-1},-15 f_d^{-1}) \approx 0.35$ points to a correlation between $V_{F3}$ and $V_{F4}$ shifted by $\theta_1 = 15 f_d^{-1}$ backwards.  $\theta_1 = 15 f_d^{-1}$ corresponds to a phase shift of $2 \pi f_1 \theta_1 = 2 \pi$, where $f_1 = 1/15 f_d \approx 0.5$ Hz. This implies that signal $V_{F4}(t)$ ``lags" from signal $V_{F3}(t)$ by phase $2\pi$. Likewise, if we consider a local maximum $q(40 f_d^{-1},15 f_d^{-1}) \approx 0.25$, we can conclude that signal $V_{F4}(t)$ goes ahead of signal $V_{F3}(t)$ by the same time interval or phase $2\pi$. Therefore, we observe a phase synchronization of signals $V_{F3}$ and $V_{F4}$, which takes place in the cortex of subject S47 within studied time interval. The presence of local maxima $q(40 f_d^{-1},-30 f_d^{-1})$ and $q(40 f_d^{-1},30 f_d^{-1})$ in Fig. \ref{fig5}c with values exceeding 0.1 implies that the phase synchronization of signals $V_{F3}$ and $V_{F4}$ occurs with a constant frequency $f_1$, which allows us to use the term ``frequency-phase synchronization". It should be noted that the observed frequency $f_1$ is individual-specific and can vary near its base value.

The threshold value of 0.1, which was marked on the plane projections of cross-correlation function $q(\tau_0,\theta)$ by a horizontal dashed line, was selected based on the condition of closest agreement between our cross-correlation analysis and medical diagnosis at NCPH. The selection of $\tau_0 = 40 f_d^{-1}$ for the cross-section of cross-correlation function $q(\tau_0,\theta)$ was done based on the observed characteristic period $\approx 15 f_d^{-1}$ in recorded signals. It can be assumed that at time lags $\tau_0$ exceeding two periods of oscillations the correlation links in stochastic components of difference moments are lost and the steady-state variance is reached \cite{Tim07a,Tim07b}.

The cross-correlation dependences $q(\tau,\theta)$ and $q(40 f_d^{-1},\theta)$ for the $F_3$ and $F_4$ signals of subject S177 also demonstrate the frequency-phase synchronization. But in contrast to S47, here, the number of local maxima exceeding 0.1 is reduced and their shape is significanly deformed. In other words, the effects of frequency-phase synchronization are less pronounced in the case of group II as compared to group I. At the same time, a comparative analysis of frequency-phase synchronization and FNS parameters with medical diagnosis still indicates that the susceptibility to schizophrenia for group II is relatively low.

A similar logic applies to the analysis of cross-correlation dependences  $q(\tau,\theta)$ and their plane projections $q(40 f_d^{-1},\theta)$ of the EEG signals for groups III and IV (575, 221) (Fig. \ref{fig6}). In general, the structure of three-dimensional plots for cross-correlation function gets largely deformed, and the number and magnitude of local maxima substantially decrease for groups with higher identification numbers (III and IV).
 
\section{Classification of EEG signals based on degree of frequency-phase synchronization}
\label{5}

The results presented above suggest that quantitative estimates characterizing the degree of frequency-phase synchronization in signals $V_{F3}(t)$ and $V_{F4}(t)$ should be introduced in addition to FNS parameters for the signals themselves in order to objectively and unambiguously estimate subjects' susceptibility to schizophrenia. To estimate the degree of synchronization, we can use the plane projection of $q(\tau, \theta)$ at value $\tau_0$ and count the number of pairs of local maxima along the $\theta$ axis, which are symmetric with respect to $\theta = 0$, in the selected range $|\theta| \le 150 f_d^{-1}$ for each subinterval with length $T = 800 f_d^{-1} = 6.25$ s. The selected ranges $|\theta| \le 150 f_d^{-1}$ with respect to $\theta = 0$ are marked by vertical dashed lines in Figs. \ref{fig5}-\ref{fig6}. The magnitudes of all counted local maxima must exceed a specific threshold value (0.1 in this case). The number of such pairs $n_s$ per unit time averaged for all 9 subintervals can be used to characterize the frequency $f_s=n_s/T$ of ``deep synchronization" for signals $V_{F3}(t)$ and $V_{F4}(t)$ and be regarded as the basic criterion in assigning a subject to a specific category. The discreteness of pair number $n_s$ implies that the values of $f_s$ are multiples of 0.16 Hz for the chosen value of $T$. Therefore, the determined values of frequency $f_s$ should only be regarded as rough estimates (``rounded" up to 0.16 Hz). At the same time, the number of observed pairs $n_s$ can be considered as an objective criterion for determining the susceptibility of a child/adolescent to schizophrenia (lower risk for developing schizophrenia in adult age) for the given value of $T$.

\begin{table}
\caption{\label{tab2}Classification of subjects into categories corresponding to different risk levels for developing schizophrenia based on the analysis of $F_3$ and $F_4$ EEG signals: I - ``healthy", II - ``conditionally healthy", III - ``susceptible to schizophrenia", IV - ``highly susceptible to schizophrenia".  (*) denotes the cases of strong nonstationarity in EEG signals}
\begin{center}
\scriptsize
\begin{tabular}{|c|@{\hspace{0.07cm}}c@{\hspace{0.07cm}}|@{\hspace{0.07cm}}c@{\hspace{0.07cm}}|c|@{\hspace{0.07cm}}c@{\hspace{0.07cm}}|@{\hspace{0.07cm}}c@{\hspace{0.07cm}}|c|@{\hspace{0.07cm}}c@{\hspace{0.07cm}}|@{\hspace{0.07cm}}c@{\hspace{0.07cm}}|c|@{\hspace{0.07cm}}c@{\hspace{0.07cm}}|@{\hspace{0.07cm}}c@{\hspace{0.07cm}}|}
    \hline
    \multicolumn{3}{|c|}{I} & \multicolumn{3}{|c|}{II} & \multicolumn{3}{|c|}{III} & \multicolumn{3}{|c|}{IV}\\
    \hline

    Subject & $f_s$, Hz & $S_s(T_{01}^{-1})$ & Subject & $f_s$, Hz & $S_s(T_{01}^{-1})$ & Subject & $f_s$, Hz & $S_s(T_{01}^{-1})$ &Subject & $f_s$, Hz & $S_s(T_{01}^{-1})$ \\
    %\hline
    $\mathstrut$ & $\mathstrut$ & $\times 10^3,$ & $\mathstrut$ & $\mathstrut$ & $\times 10^3,$ & $\mathstrut$ & $\mathstrut$ & $\times 10^3,$ & $\mathstrut$ & $\mathstrut$ & $\times 10^3,$ \\
    $\mathstrut$ & $\mathstrut$ & $\mu$V$^2/f_d$ & $\mathstrut$ & $\mathstrut$ & $\mu$V$^2/f_d$ & $\mathstrut$ & $\mathstrut$ & $\mu$V$^2/f_d$ & $\mathstrut$ & $\mathstrut$ & $\mu$V$^2/f_d$ \\
    \hline
\textbf{S94} &0.48&0.26& \textbf{416} &0.32&0.01& \textbf{S12} &0.16&0.04& \textbf{113} & $<0.01$ & 575 \\ 
\textbf{454} &0.48&0.38& \textbf{S42} &0.32&0.26& \textbf{586} &0.16&0.08& \textbf{221} & $<0.01$ & 658 \\ 
\textbf{S182} &0.48&0.68& \textbf{425} &0.32&0.28& \textbf{S158} &0.16&0.1& \textbf{307} &0.16& 7.2 \\ 
\textbf{S165} &0.48&0.8& \textbf{S154} &0.32&0.54& \textbf{84} &0.16&0.13& \textbf{S18} &0.16& 7.2 \\ 
\textbf{S20} &0.48&1.67& \textbf{S50} &0.32&0.55& \textbf{83} &0.16&0.13& \textbf{719} &0.16& 9.5 \\ 
\textbf{S60} &0.48&2.8& \textbf{S53} &0.32&0.6& \textbf{S72} &0.16&0.16& \textbf{387\_03} &0.16& 11.6 \\ 
\textbf{S163} &0.64&0.06& \textbf{S152} &0.32&0.8& \textbf{517} &0.16&0.2& \textbf{314} &0.16& 13.3 \\ 
\textbf{S78} &0.64&0.23& \textbf{S179} &0.32&0.89& \textbf{401} &0.16&0.27& \textbf{509} &0.16& 16.9 \\ 
\textbf{S155} &0.8&1.26& \textbf{S10} &0.32&0.91& \textbf{249} &0.16&0.48& \textbf{219} &0.16& 17.5 \\ 
\textbf{S164} &0.8&4& \textbf{S178} &0.32&1.02& \textbf{S176} &0.16&0.6& \textbf{33} &0.16& 20.5 \\ 
\textbf{S85} &0.96&0.06& \textbf{88} &0.32&2.1& \textbf{485} &0.16&0.74& \textbf{155} &0.16& 21.1 \\ 
\textbf{S153} &1.28&0.04& \textbf{S157} &0.32&5.4& \textbf{S169} &0.16&0.8& \textbf{423} &0.16& 41.3 \\ 
\textbf{S47} &1.28&5.7& \textbf{S27} &0.32&5.88& \textbf{S174} &0.16&1.26& \textbf{22} &0.16& 50 \\ 
\textbf{S173} &1.6&2.4& \textbf{S55} &0.48&5.1& \textbf{229} &0.16&1.72& \textbf{683} &0.16& 55.8 \\ 
\textbf{276*} &0.48&0.56& \textbf{156} &0.48&26.9& \textbf{382} &0.16&2.3& \textbf{515} &0.16& 138 \\ 
\textbf{342*} &0.64&0.01& \textbf{S26} &0.64&33.2& \textbf{32} &0.16&3.6& \textbf{573} &0.16& 169 \\ 
\textbf{312*} &0.64&0.22& \textbf{351*} &0.32&0.27& \textbf{192} &0.16&4& \textbf{S167} &0.16& 226 \\ 
$\mathstrut$ & $\mathstrut$ & $\mathstrut$ & \textbf{S170*} &0.32&0.12& \textbf{103} &0.16&4.2& \textbf{642} &0.32& 336 \\ 
$\mathstrut$ & $\mathstrut$ & $\mathstrut$ & \textbf{S177*} &0.48&0.16& \textbf{575} &0.16&6.2& \textbf{508} &0.32& 349 \\ 
$\mathstrut$ & $\mathstrut$ & $\mathstrut$ & \textbf{S59*} &0.48&0.33& \textbf{585} &0.32&8.3& $\mathstrut$ & $\mathstrut$ & $\mathstrut$ \\ 
$\mathstrut$ & $\mathstrut$ & $\mathstrut$ & \textbf{S43*} &0.48&41.8& \textbf{548} &0.32&11.6& $\mathstrut$ & $\mathstrut$ & $\mathstrut$ \\ 
$\mathstrut$ & $\mathstrut$ & $\mathstrut$ & $\mathstrut$ & $\mathstrut$ & $\mathstrut$ & \textbf{540} &0.32&18.5& $\mathstrut$ & $\mathstrut$ & $\mathstrut$ \\ 
$\mathstrut$ & $\mathstrut$ & $\mathstrut$ & $\mathstrut$ & $\mathstrut$ & $\mathstrut$ & \textbf{S31} &0.32&27.2& $\mathstrut$ & $\mathstrut$ & $\mathstrut$ \\ 
$\mathstrut$ & $\mathstrut$ & $\mathstrut$ & $\mathstrut$ & $\mathstrut$ & $\mathstrut$ & \textbf{S196} &0.32&27.6& $\mathstrut$ & $\mathstrut$ & $\mathstrut$ \\ 
$\mathstrut$ & $\mathstrut$ & $\mathstrut$ & $\mathstrut$ & $\mathstrut$ & $\mathstrut$ & \textbf{429*} &0.32&1.83& $\mathstrut$ & $\mathstrut$ & $\mathstrut$ \\
$\mathstrut$ & $\mathstrut$ & $\mathstrut$ & $\mathstrut$ & $\mathstrut$ & $\mathstrut$ & \textbf{510*} &0.16&1& $\mathstrut$ & $\mathstrut$ & $\mathstrut$ \\
$\mathstrut$ & $\mathstrut$ & $\mathstrut$ & $\mathstrut$ & $\mathstrut$ & $\mathstrut$ & \textbf{387\_02*} &0.16&1.01& $\mathstrut$ & $\mathstrut$ & $\mathstrut$ \\
    \hline
  \end{tabular}
\normalsize
\end{center}
\end{table}

Our analysis showed that for 23 subjects out of total 84 the frequency $f_s$ was at least 0.48 Hz. For 2 subjects (S47 and S153) this frequency was 1.28 Hz, and for subject S173 it reached the value of 1.6 Hz. These 3 values significantly exceed the average $f_s$ for the group of subjects labeled as ``healthy" by NCPH. These anomalies in the level of frequency-phase synchronization require an additional analysis and may be related to possible precursors of a hidden disorder. There are known examples in medicine when the increase of synchronization level in the dynamics of specific cortex areas beyond some threshold value is considered as an indicator of a disorder. An example of an anomalously high activity in neural ensembles associated with an epileptic seizure is discussed in Refs. \cite{Pan2010,Tim10b}. It is possible that a similar event took place in subject S47 on some of the considered time subintervals. As the authors do not have any information on the later examinations of these 3 subjects labeled as ``healthy" by NCPH, we will assign them to category I (``healthy") in the classification described below. For 23 subjects, the $f_s$ parameter was equal to 0.32 Hz; for 36, $f_s$ = 0.16 Hz; and for the two remaining subjects, $f_s < 0.01$ Hz. The lower values of $f_s$ generally corresponded to a higher risk for developing schizophrenia (determined by medical diagnosis at NCPH).

A comparative analysis of the values of  $S_s(T_{01}^{-1})$ (maximum value for signals $F_3$ and $F_4$) for the EEG signals belonging to different categories suggested that the spikiness factor can be used together with $f_s$ as a basic diagnostic sign for schizophrenia-spectrum disorders. The higher values of $S_s(T_{01}^{-1})$ and lower values of $f_s$ correspond to higher risks for developing schizophrenia. As a result, the classification presented in Table \ref{tab2} was built using the optimality criterion for the values of these two parameters that leads to the minimal discrepancy between the results of medical diagnosis and proposed partitioning. The limiting cases of the partitioning are: (1) large values of parameter $f_s$ accompanied by small values of parameter $S_s(T_{01}^{-1})$ correspond to a healthy subject; (2) small values of $f_s$ with large values of $S_s(T_{01}^{-1})$ correspond to a high risk for developing schizophrenia. In contradictory cases, such as S26, S43, and 156, other FNS parameters were also used to assign the subjects to appropriate groups. For example, the high levels of ``deep synchronization" ($f_s \thicksim 0.5$) and stochasticity in subjects  S26, S43, and 156 allowed us to conclude that the subjects belong to group II.

The complete algorithm for partitioning subjects into 4 groups corresponding to different risk levels for developing schizophrenia is written as:

\begin{enumerate}[(1)]
\item Group I. Necessary condition: $f_s \ge 0.48$ Hz. If $f_s = 0.48$ Hz and $T_1 < T$, the following condition should also hold: $S_s(T_{01}^{-1}) < 3 \times 10^3 \mu$V$^2 f_d^{-1}$.  If $f_s = 0.48$ Hz and $T_1 \ge T$ (nonstationary), the following condition should also hold: $S_s(T_{01}^{-1}) < 6 \times 10^2 \mu$V$^2 f_d^{-1}$. If $f_s$ = 0.64 Hz,  the following condition should also hold: $S_s(T_{01}^{-1}) < 3 \times 10^2 \mu$V$^2 f_d^{-1}$.
\item Group II. Necessary condition: $0.32 \le f_s \le 0.64$ Hz. If $f_s = 0.32$ Hz and $T_1 < T$, the following condition should also hold: $S_s(T_{01}^{-1}) < 6 \times 10^3 \mu$V$^2 f_d^{-1}$. If $f_s = 0.32$ Hz and $T_1 \ge T$ (nonstationary), the following condition should also be true:  $S_s(T_{01}^{-1}) < 3 \times 10^2 \mu$V$^2 f_d^{-1}$. If $f_s = 0.48$ Hz and $T_1 < T$, the following condition should also hold: $3 \times 10^3 \mu$V$^2 f_d^{-1} \le S_s(T_{01}^{-1})  < 5 \times 10^4 \mu$V$^2 f_d^{-1}$.  If $f_s = 0.48$ Hz and $T_1 \ge T$ (nonstationary), the following condition should also hold: $6 \times 10^2 \mu$V$^2 f_d^{-1} \le S_s(T_{01}^{-1}) < 5 \times 10^4 \mu$V$^2 f_d^{-1}$.  If $f_s$ = 0.64 Hz,  the following condition should also hold: $S_s(T_{01}^{-1}) \ge 3 \times 10^2 \mu$V$^2 f_d^{-1}$.
\item Group III. Necessary condition: $0.16 \le f_s \le 0.32$ Hz.  If $f_s = 0.16$ Hz and $T_1 < T$, the following condition should also hold: $S_s(T_{01}^{-1}) < 7 \times 10^3 \mu$V$^2 f_d^{-1}$. If $f_s = 0.16$ Hz and $T_1 \ge T$ (nonstationary), the following condition should also be true:  $S_s(T_{01}^{-1}) < 2 \times 10^3 \mu$V$^2 f_d^{-1}$. If $f_s = 0.32$ Hz and $T_1 < T$, the following condition should also hold: $ 6 \times 10^3 \mu$V$^2 f_d^{-1} \le S_s(T_{01}^{-1}) <  3 \times 10^4 \mu$V$^2 f_d^{-1}$. If $f_s = 0.32$ Hz and $T_1 \ge T$ (nonstationary), the following condition should also be true:  $3 \times 10^2 \mu$V$^2 f_d^{-1} \le S_s(T_{01}^{-1}) < 3 \times 10^3 \mu$V$^2 f_d^{-1}$.
\item Group IV. Necessary condition: $f_s \le 0.32$ Hz. If $f_s = 0.16$ Hz, the following condition should also hold: $S_s(T_{01}^{-1}) \ge 7 \times 10^3 \mu$V$^2 f_d^{-1}$. If $f_s = 0.32$ Hz, the following condition should also hold: $S_s(T_{01}^{-1}) \ge 3 \times 10^4 \mu$V$^2 f_d^{-1}$.
\end{enumerate}

It can be seen from Table \ref{tab2} that the difference between this partitioning of 84 subjects and medical diagnoses made by interviewing the subjects is 20$\%$, given that the subjects in groups I and II are mapped to the ``healthy" group of the medical diagnosis and groups II and IV to the ``susceptible to schizophrenia" group.

It is known that the signals produced \emph{in vivo} by a human organism contain both a collection of low-frequency (``resonance") components and higher-frequency stochastic variability \cite{Iva99,Pen93,Gol02,Pen95,Boc00,Yul05,Yul06}. We believe that the random components are responsible for the adjustment of the organism to constantly changing external conditions, which also manifests itself as the rearrangement of resonance components. Our results imply that this general feature of \emph{in vivo} systems can be seen in the analysis of frequency-phase synchronization. Our study shows that the degree of disruption in frequency-phase synchronization can be considered as a diagnostic sign in identifying possible pathological changes in the higher neural activity of children/adolescents. To quantify frequency phase-synchronization and its disruption, we introduced a parameter $f_s$, specific frequency characterizing the synchronization depth, and FNS parameters for stochastic signal components, respectively. It can be assumed that parameter $\sigma$, which varied within the same ranges for subjects in groups I-III and had slightly higher values for group IV, is a measure of the randomness responsible for the adjustment of the organism to changing external conditions. On the other hand, the spikiness factor $S_s(T_{01}^{-1})$, which varied in the analyzed signals up to 4-5 orders of magnitude and was highest for subjects in group IV, appears to be related not just to the stochasticity associated with jump- and spike-like irregularities, but also to the disruption of frequency-phase synchronization.

\section{Concluding remarks}
\label{6}

Our analysis demonstrates a high specificity of EEG signals and the processes of their space-time synchronization, which may be attributed to an individual-specific set of natural frequencies and their interferential contributions rearranging in time. These rearrangement processes in the brain are most likely associated with the organism's response to the action of external factors. The specificity of EEG signals is also observed in the behavior of stochastic components, particularly, in the values of FNS parameter $\sigma$, the 
 standard deviation of the value of measured dynamic variable from the slowly varying resonant (regular) component, and in the flicker-noise dependences of stochastic components. This high level of specificity of biomedical signals makes the diagnosis of functional disruptions in organism's subsystems complex.

We show that in the case of schizophrenia one can substantially reduce the effect of specific features of biomedical signals by analyzing the quantitative measures of frequency-phase synchronization between the EEG signals recorded at different cortex areas. According to our working hypothesis, the occurrence of a psychiatric pathology leads to a significant deviation of the frequency-phase synchronization level from it is normal value. As the minimum level of disruptions under normal conditions may be individual-specific, the estimation of ``individual norms" should become a topic of high concern. This topic is related to a more general problem of developing the principles for personalized medicine, an emerging medical model emphasizing the customization of healthcare.

We believe that the proposed working hypothesis can be applied to the diagnosis of other psychiatric disorders as well as neurodegenerative disorders (Parkinson's, Alzheimer's, and Hungtington's diseases; progressive supranuclear palsy; and epilepsy). This implies that the diagnosis of any of the above disorders requires at least two digitized biomedical signals measured in different cortex areas at the same time. In the presented study, we analyzed the EEG signals from scalp-mounted electrodes at locations $F_3$ and $F_4$ for 84 children/adolescents and were able to classify them into 4 groups based on frequency-phase synchronization levels, which appear to be related (judging from the comparison with medical diagnoses) to the subjects' risk levels for developing schizophrenia in adult age. For other psychiatric and neurodegenerative disorders, appropriate cortex sites should be selected based on preliminary analysis or available literature. The cross-correlation analysis presented above can also be used to estimate the effectiveness of a therapeutic method by comparing the values of frequency-phase synchronization levels before and after the therapy.

\section*{Acknowledgements}
The authors are grateful to A. F. Iznak, N. L. Gorbachevskiy, and I. A. Kozlova at the National Center for Psychiatric Health of the Russian Academy of Medical Sciences for the EEG signal recordings analyzed in this study. This work was partially supported by the Russian Foundation for Basic Research, grants nos. 09-04-12094-ofi\_m, 10-02-01346-a, 11-02-00540-a; Russian Humanities Scientific Fund, grant no. 09-06-00671-a.

%% The Appendices part is started with the command appendix;
\appendix
 
\section{Parameterization algorithm in discrete form}
\label{A}

 Consider a time series $V_d(k)$. The subscript $d$ here and below is used to denote the discrete form of expressions. Let $N_t$ be the number of points corresponding to the selected averaging interval $T$, $M$  be the number of 
 points used in estimating the autocorrelation function. In this case, the parameterization procedure can be written as follows:
\newline 1. Calculate the arithmetic mean for the signal:
%eq6
\begin{equation}
\mu _V  = {1 \over {N_t }}\sum\limits_{k = 1}^{N_t} {V_d\left( k \right)}. \label{eqA7}
\end{equation}
\newline 2. Subtract the arithmetic mean from the series $V_d(k)$:
\begin{equation}
\mathop {V_d}\limits^-  \left( k \right) = V_d\left( k \right) - \mu _V. \label{eqA8}
\end{equation}
\newline 3. Calculate the autocorrelation function for the series $\mathop {V_d}\limits^ -$:
\begin{equation}
 \psi_d (p) = {1 \over {N_t  - p}}\sum\limits_{k = 1}^{N_t  - p} {\mathop {V_d}\limits^ -  \left( k \right)\,\mathop {V_d}\limits^ -  \left( {k + p} 
\right)},   p = 0..M.  \label{eqA9}
\end{equation}
Let the autocorrelation interval $M$ be $T/4$ (higher values of $M$ will result in 
the loss of statistical information in estimating the autocorrelation function). To go from discrete form to the continuous one, one can use the following expression: $p = N_t \tau/T$.
\newline 4. Calculate the discrete cosine transform of the autocorrelation function:
\begin{equation}
 S_d(q) = \psi_d (0) + \psi_d \left( M \right)\left( { - 1} \right)^q  + 2\sum\limits_{p = 1}^{M - 1} {\psi_d (p)\cos \left( {{{\pi  \,q\,p} \over 
M}} \right)}, \label{eqA10}
\end{equation}
 where $q=0..M$. For $q=1..M-1$, $S_d(q)$ should be multiplied by 2, which is the standard procedure for discrete Fourier 
transforms to take into account the spectral values in the second half of the frequency range. Here, relations $q=2 f f_d^{-1} M$ and $S_d(q)=S(f)\times f_d$  describe the equivalence between the discrete and continuous 
 forms of power spectrum estimate.
 \newline 5. Calculate $S_{sd}(0)$  as the average value of the power spectrum for the points 1 and 2 (point 0, which corresponds to the zero 
frequency, is not used in calculating $S_{sd}(0)$):
\begin{equation}
S_{sd} \left( 0 \right) = {{S_d\left( 1 \right) + S_d\left( 2 \right)} \over 2}. \label{eqA11}
\end{equation}
\newline 6. Interpolate $|S_d(q)|$ given by Eq.~(\ref{eqA10}) using the expression:
\begin{equation}
S_{sd} (q) = {{S_{sd} (0)} \over {1 + (\pi {q \over M}T_{01d} )^n }} \label{eqA12}
\end{equation}
 by the method of nonlinear least-square fitting to determine the values of parameters $n$ and $T_{01d}$. The fitting is done on the basis of a 
 double logarithmic scale, dividing the entire series into a set of equal intervals. We used the trust-region algorithm for nonlinear square 
fitting, which is built in MATLAB v.7 or higher \cite{Bra99}.
\newline 7. Separate out the resonant component:
\begin{equation}
S_{rd} \left( q \right) = S_d\left( q \right) - S_{sd} \left( q \right), q = 0..M. \label{eqA13}
\end{equation}
 \newline 8. Calculate the autocorrelation function for the resonant component as the inverse discrete cosine transform of $S_{rd}(q)$. When 
 $q=1..M-1$, divide  $S_{rd}(q)$ by 2 to take into account the spectral values in the second half of the frequency range. Then calculate the 
inverse cosine transform:
\begin{eqnarray}
\psi_{rd} (p) = {1 \over {2M}} \left\{ {S_{rd} (0) + S_{rd} \left( M \right)\left( { - 1} \right)^p} \right\} \nonumber \\
+ {1 \over {2M}} \left\{ {2\sum\limits_{q = 1}^{M - 1} {S_{rd} (q)\cos \left( {{{\pi  \,p\,q} \over M}} \right) } } \right\}. \label{eqA14}
\end{eqnarray}
9. Calculate the difference moment for the resonant component:
\begin{equation}
\Phi_{rd}^{(2)} (p) = 2\left[ {\psi_{rd} (0) - \psi_{rd}  (p)} \right], p = 0..M. \label{eqA15}
\end{equation}
The continuous equivalent of $\Phi_{rd}^{(2)} (p)$ is $\Phi_{r}^{(2)} (\tau)$.
\newline 10. Calculate the difference moment for the experimental series:
\begin{equation}
 \Phi_d^{(2)} (p\,) = {1 \over {N_t  - p}}\sum\limits_{k = 1}^{N_t  - p} {\left[ {\mathop {V_d}\limits^ -  (k) - \mathop {V_d}\limits^ -  (k + p)} 
\right]^2 }. \label{eqA16}
\end{equation}
\newline 11. Calculate the difference moment for the random component:
\begin{equation}
\Phi _{esd}^{(2)} (p) = \Phi_d ^{\left( 2 \right)} (p) - \Phi _{rd}^{(2)} (p). \label{eqA17}
\end{equation}
The continuous equivalent of $\Phi_{esd}^{(2)} (p)$ is $\Phi_{es}^{(2)} (\tau)$.
 \newline 12. Determine the parameters $\sigma$, $H_1$, $T_{1d}$ by fitting $\Phi _{esd}^{(2)} (p)$ in Eq.~(\ref{eqA17}) to the interpolation 
expression of the anomalous diffusion type\cite{Tim10a}:
\begin{equation}
 \Phi_{sd}^{(2)} (p) = 2\sigma ^2  \times \left[ {1 - \Gamma ^{ - 1} (H_1 ) \times \Gamma (H_1 ,p/T_{1d} )} \right]^{2}, \label{eqA18}
\end{equation}
 where $\Gamma (s,x) = \int\limits_x^\infty  {\exp ( - t) t^{s - 1} dt},\, \Gamma (s) = \Gamma (s,0)$, using the same least-square fitting 
method as in step 6.
 \newline 13. Calculate $S_{sd}(T_{01d}^{-1})$ by Eq.~(\ref{eqA12}).
 \newline 14. After the values of all six FNS parameters - $\sigma$, $T_{01d}$, $T_{1d}$, $H_1$, $n$, $S_{sd}(T_{01d}^{-1})$ - are determined, calculate the dimensional values for $T_{01d}$, $T_{1d}$,  $S_{sd}(T_{01d}^{-1})$: $T_0 = T_{01d} \times \Delta t$, $T_1 = T_{1d} \times \Delta t$, $S_s(T_{01}^{-1}) = S_{sd}(T_{01d}^{-1}) \times \Delta t$, where $\Delta t = f_d^{-1}$ .
\newline 15. Calculate the relative error $\epsilon_\Phi$ in the interpolation of difference moment $\Phi_d^{(2)} (p)$:
\begin{equation}
 \epsilon _\Phi   = {{\sum\limits_{p = 1}^M {\left| {\Phi_d ^{(2)} \left( p \right) - \Phi _{rd}^{(2)} \left( p \right) - \Phi _{sd}^{(2)} \left( p 
\right)} \right|} } \over {\sum\limits_{p = 1}^M {\Phi_d ^{(2)} \left( p \right)} }} \times 100\%.  \label{eqA19}
\end{equation}
 Here, the error is determined as the ratio of the difference of areas between the experimental structure function and the total 
 interpolation function to the area of the experimental structure function. The areas are calculated by numerical integration using the 
rectangle method because the original series have a rather large number of points. The parameterization is successful if 
$\epsilon _\Phi   \le 10\% $ \cite{Tim10a}.

\section{Cross-correlation function in discrete form}
\label{B}
In discrete form, Eq.~(\ref{eq17}) is written as
\begin{equation}
 q_{ijd} (n_\tau,n_\theta  ) = {{\sum\limits_{k = U\left[ { - n_\theta  } \right]\,\left| {n_\theta  } \right| + 1}^{N_t  - n_\tau   - 
 U\left[ {n_\theta  } \right]\,\left| {n_\theta  } \right|} {\left[ {V_{id} (k) - V_{id} (k + n_\tau  )} \right]\,\,\left[ {V_{jd} (k + n_\theta  ) - 
 V_{jd} (k + n_\theta   + n_\tau  )} \right]\,} } \over {\sqrt {\sum\limits_{k = U\left[ { - n_\theta  } \right]\,\left| {n_\theta  } \right| + 
 1}^{N_t  - n_\tau   - U\left[ {n_\theta  } \right]\,\left| {n_\theta  } \right|} {\left[ {V_{id} (k) - V_{id} (k + n_\tau  )} \right]^2 \,} } 
 \,\,\sqrt {\sum\limits_{k = U\left[ { - n_\theta  } \right]\,\left| {n_\theta  } \right| + 1 + n_\theta  }^{N_t  - n_\tau   - U\left[ 
{n_\theta  } \right]\,\left| {n_\theta  } \right| + n_\theta  } {\left[ {V_{jd} (k) - V_{jd} (k + n_\tau  )} \right]^2 \,} } }},  \label{eqB1}
\end{equation}
where $n_\tau   = \left\lfloor {\tau /\Delta t} \right\rfloor$, $n_\theta   = \left\lfloor {\theta _{ij} /\Delta t} \right\rfloor$, $
U\left[ x \right] = \left\{ \matrix{
  1,\,\,\,x \ge 0; \hfill \cr 
  0,\,\,\,x < 0. \hfill \cr}  \right.
$

%% References


\bibliographystyle{model1-num-names}
\bibliography{EEG-PhysicaA}






% \end{thebibliography}

\end{document}